\documentclass[aps,prl,twocolumn,groupedaddress,notitlepage,showpacs,floatfix,superscriptaddress]{revtex4-1}
\pdfoutput=1

\usepackage{graphicx,graphics,epsfig,subfigure,times,bm,bbm,amssymb,amsmath,amsthm,mathrsfs,MnSymbol}
\usepackage{gensymb}
\usepackage{amsfonts}
\usepackage[matrix,frame,arrow]{xypic}
\usepackage[pdfstartview=FitH]{hyperref}
\hypersetup{
    colorlinks=true,       
    linkcolor=red,          
   citecolor=magenta,        
    filecolor=magenta,      
    urlcolor=cyan,           
    runcolor=cyan
}
\usepackage[pdftex]{color}
\usepackage{braket}
\usepackage{enumerate}
\usepackage[normalem]{ulem}
\usepackage[usenames,dvipsnames]{xcolor}
\usepackage{multirow}
\usepackage{mathtools}
\usepackage{ulem}

\definecolor{orange}{rgb}{1,0.5,0}

\newtheorem{definition}{Definition}
\newtheorem{proposition}{Proposition}

\newcommand{\ignore}[1]{}

\ignore{
\documentclass[eprintnumbers,amsmath,amssymb,onecolumn,a4paper,caption ]{article}
\usepackage{amsfonts}
\usepackage{amssymb}
\usepackage{mathrsfs}
\usepackage{mathbbold}
\usepackage{bbm}
\usepackage{mathrsfs}
\usepackage{dcolumn}
\usepackage{bm}
\usepackage{times,epsfig,amssymb,amsmath}
\usepackage{float}
\usepackage{subfigure}

\usepackage{color}

}

\begin{document}

\title{Bounded light cone and robust topological order out of equilibrium}
\author{Yu~Zeng}\thanks{zengyunk@gmail.com}
\affiliation{Beijing National Laboratory for Condensed Matter Physics, Institute of Physics, Chinese Academy of Sciences, Beijing 100190, China}
\affiliation{Shandong Inspur Intelligence Research Institute Co., Ltd, Jinan 250100, China}
\author{Alioscia~Hamma}\thanks{Alioscia.Hamma@umb.edu}
\affiliation{Department of Physics, University of Massachusetts Boston, 100 Morrissey Blvd, Boston MA 02125, USA}
\affiliation{Dipartimento di Fisica `Ettore Pancini', Universit\`a degli Studi di Napoli Federico II, Via Cintia 80126, Napoli, Italy}
\author{Yu-Ran Zhang}
\affiliation{Theoretical Quantum Physics Laboratory, RIKEN Cluster for Pioneering Research, Wako-shi, Saitama 351-0198, Japan}
\affiliation{School of Physics and Optoelectronics, South China University of Technology, Guangzhou 510640, China}
\author{Jun-Peng Cao}
\affiliation{Beijing National Laboratory for Condensed Matter Physics, Institute of Physics, Chinese Academy of Sciences, Beijing 100190, China}
\author{Heng~Fan}\thanks{hfan@iphy.ac.cn}
\affiliation{Beijing National Laboratory for Condensed Matter Physics, Institute of Physics, Chinese Academy of Sciences, Beijing 100190, China}
\author{Wu-Ming Liu}\thanks{wliu@iphy.ac.cn}
\affiliation{Beijing National Laboratory for Condensed Matter Physics, Institute of Physics, Chinese Academy of Sciences, Beijing 100190, China}

\begin{abstract}
The ground state degeneracy of topologically ordered gapped Hamiltonians is the bedrock for self-correcting quantum memories, which are unfortunately not stable away from equilibrium even at zero temperature. This plague precludes practical robust self-correction since stability at zero temperature is a prerequisite for finite-temperature robustness. In this work, we show that the emergence of a bounded light cone renders the unitary time evolution a quasi-adiabatic continuation that preserves topological order, with the initial ground space retaining its macroscopic distance at all times as a quantum code. We also show how bounded light cones can emerge through suitable perturbations in Kitaev's toric code and honeycomb model. Our results suggest that topological orders and self-correcting quantum memories can be dynamically robust at zero temperature.
\end{abstract}

\maketitle





{\em Introduction}. The gapped quantum phases of matter with topological order (TO) go beyond the Landau paradigm and possess locally indistinguishable degenerate ground states on closed space manifolds \cite{wenbook,Wen1990}, whose properties make them promising candidates as \emph{self-correcting} quantum memories (QMs) \cite{dennis,kitaev:2003}. Correspondingly,  quantum error-correcting codes (QECs) provide toy models of TO with Hamiltonians that are the sum of commuting local projectors and  ground spaces as code spaces with \emph{macroscopic distance} \cite{kitaev:2003}. Along with decades of theoretical exploration, the intimacy between TO and quantum information processing \cite{nayak:2008} has recently sparked much experimental effort to realize topologically ordered states \cite{Satzinger2021,Semeghini2021} and QECs \cite{Goole2021,Egan2021,Krinner2022,Zhao2022,Google2023,Bluvstein2023}.

The code space with TO is robust in the sense that the gap and the topological degeneracy are stable against inevitable small local perturbations \cite{Bravyi2010a}; meanwhile, the states in the same phase are connected by a \emph{quasi-adiabatic continuation} that preserves the macroscopic distance \cite{Bravyi2010a,HastingsWen2005,Osborne2007}. However, this `robustness' is too readily adopted: the initially prepared state generally cannot be an eigenstate of the perturbed Hamiltonian, so the non-equilibrium effects of quantum dynamics must be considered \cite{rmp2016}. Unlike active error correction, self-correcting QMs need no constant error correction. Consider the scenario: the QM is well-isolated, and thermal noise is introduced when external apparatuses (applying logical operations or readouts) couple to the system. The dynamics of the encoded state, dominated by the system's perturbed Hamiltonian, are approximately unitary until the apparatuses intervene. The QM retains macroscopic distance during its lifetime, and error correction is performed after the coupling. Although we hope the QM and its TO have exponentially long lifetimes away from equilibrium, unfortunately, they generally do not \cite{Kay2009,Pastawski2010,kay2011,Bravyi2012,Tsomokos2009,Yu2016,disorderTCM1,disorderTCM2}. 

In this letter, we show that the presence of a bounded light cone (BLC) renders the unitary time evolution a quasi-adiabatic continuation, where the BLC emerges from the Lieb-Robinson bound (LRB); 
and the TO quantum phase remains well-defined under unitary time evolution, with the initial ground space continuing to be a QEC with macroscopic distance.  We substantialize this setting by introducing randomness in Kitaev's toric code and honeycomb model, thereby obtaining \emph{dynamical localization} with BLCs. We also present numerical results for typical non-local TO parameters to support our conclusions.	

{\em BLC and  TO.}  Consider a quantum system defined on a $D$ dimensional lattice $\Lambda$,  a metric space of sites. For $i,j\in\Lambda$, denote by $\operatorname{dist}(i, j)$ the length of the shortest path connecting $i$ to $j$. The linear size of the lattice is $L$. Define the diameter of a subset $X\subset\Lambda$ as $\operatorname{diam}\left(X\right)=\operatorname{max}_{i, j\in X}\operatorname{dist}\left(i, j\right)$ and the distance between two subsets as $\operatorname{dist}\left(X, Y\right)=\operatorname{min}_{i\in X, j\in Y}\operatorname{dist}\left(i, j\right)$. The Hilbert space is a tensor product of the local Hilbert spaces on lattice sites, $\mathscr{H}=\bigotimes_{i\in\Lambda}\mathscr{H}_{i}$, dim$(\mathscr{H}_{i})=\mathcal{O}(1)$. The Hamiltonian is $H_0=\sum_{Z}H_Z$, where $H_Z$ is a bounded operator supported on a bounded set $Z\subset\Lambda$. The dimension of the ground-space projector $P_0$ depends only on the topology of the space manifold. TO is defined by the existence of a length $L^\ast=\Omega(L)$ such that arbitrary operator $O$, supported on a set whose diameter $\le L^\ast$, does satisfy $P_0OP_0\propto P_0$ \cite{Bravyi2010a}. 
We then say that $P_0$ is a QEC with \emph{macroscopic distance} \cite{Bravyi2010b}. Notice that a ground space $P$ is a QEC correcting error $\mathcal{E}$ \emph{iff} $PE^\dagger EP\propto P$ for all $E\in\mathcal{E}$, where $\mathcal{E}$ is a linear space of errors (\emph{quantum error-correction conditions}
\cite{NielsenChuang,Gottesman2009}). Introducing small local perturbations, we can define a continuous family of Hamiltonians, $H_s=H_0+sV$, $0\leq s\leq 1$, where the gap of $H_s$ is not closed and degeneracy splitting is exponentially small with $L$. Consequently, the ground-space projectors in the same phase are quasi-adiabatically connected by a local unitary, $P_s=U_sP_0U_s^\dagger$ \cite{Bravyi2010a,HastingsWen2005,Osborne2007}. $P_s$ is of `perturbed' TO quantified as \cite{liebrobinson2}:
\begin{definition}
$P$ is topologically ordered, if there exist $L^\ast=\Omega(L)$, $\epsilon=\operatorname{exp}(-\Omega(L^a))$ with $a>0$ and a scalar $z$, such that for any operator $O$ supported on the set whose diameter is smaller than $L^\ast$, 
\begin{eqnarray}
  \parallel POP-zP\parallel\leq\epsilon.
\end{eqnarray}
We say $P$ is of TO to accuracy $(L^\ast,\epsilon)$.
\end{definition}
For example, if $P_0$ is of TO to accuracy $(L/2,0)$ then one can prove that $P_1$ is of TO to accuracy $(L/4,e^{-\Omega(L^a)})$ \cite{Bravyi2010a,HastingsWen2005,Osborne2007}. So $P_1$ continues to be a QEC approximately with macroscopic distance for large $L$. We will show that this continuation can also appear in quantum dynamics.

\begin{figure}
\centering
\includegraphics[width=0.4\textwidth]{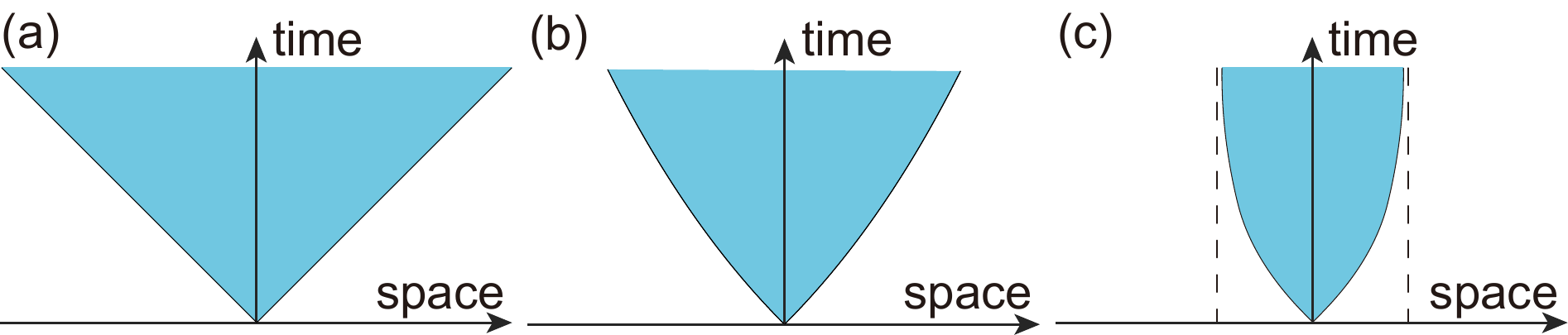}\\
\caption{ Schematic illustration of (a) linear, (b) logarithmic, and (c) bounded effective light cone. The blue regions dipict causal regions.}\label{lightcone}
\end{figure}

In nonrelativistic quantum systems with local interactions, effective light cones emerge from LRB: for any two operators $A_X$ and $B_Y$ supported on subset $X$ and $Y$ in $\Lambda$ with $\operatorname{dist}(X, Y)=l$, if $l-vt>0$, then $\|[A_{X}(t),B_{Y}]\|\leq\operatorname{exp}(-\Omega(l))$. Here, the Lieb-Robinson velocity $v$ characterizes the maximum velocity of signals in the model \cite{liebrobinson0,liebrobinson1,liebrobinson2,liebrobinson3}.
The linear Lieb-Robinson bound (LRB) underpins several key theorems in many-body physics \cite{Hastings2004a,Hastings2004b,Hastings2006,Nachtergaele2006,Hastings2015,Hastings2010}. Recent research indicates that many-body localization is notably linked to a \emph{logarithmic} light cone \cite{Kim2014,Deng2017,Chen2016,Huang2017,Chen2017,Fan2017,Swingle2017}, a concept first proven in the one-dimensional XX model with Anderson localization \cite{Burrell2007}. It was further demonstrated that the bound is actually time independent \cite{Hamza2012}, and later extended to the XY model \cite{mathlocalizationXY1,mathlocalizationXY2,XYlocalization}.
Next, we give a formal definition of the BLC and show its consequences. The three types of light cones are illustrated schematically in Fig. \ref{lightcone}.
\begin{definition}
A quantum system possesses a BLC, if any two operators $A_X$ and $B_Y$ supported on subsets $X$ and $Y$ at a distance of $\operatorname{dist}(X,Y)$ satisfy
\begin{eqnarray}\label{boundedLC}
\left\|\left[\!A_{X}(t) , B_{Y}\!\right]\right\|\!\leqslant\! C\left|X\right|\!\left\| A_{X}\right\|\!\left\|B_{Y}\!\right\|e^{-\mu\left(\operatorname{dist}(X, Y)\right)}.
\end{eqnarray}
Here $C$ and $\mu$ are nonnegative constant, $\left\|\cdots\right\|$ denotes operator norm, $\left|\cdots\right|$ the cardinality of the set.
\end{definition}
\begin{proposition}\label{P1}
If $P_0$ possesses TO to accuracy $(L^\ast,\epsilon)$, $\epsilon=\operatorname{exp}(-\Omega(L))$ and  the system with Hamiltonian $H_1$ has a BLC, then $P(t)=U(t)P_0U(t)^\dagger$ is topological ordered, where $U(t)=\operatorname{exp}(-itH_1)$.
\end{proposition}
{\begin{proof}
For any $O_A$, $\|O_A\|=1$ without loss of generality, supported on $A$ with $\operatorname{diam}(A)\leq L^\ast/2$, we will prove that there exists a scalar $z$ such that $\|P(t)O_AP(t)-zP(t)\|=\|P_0O_A(t)P_0-zP_0\|\leq\epsilon^\prime$.
Notice that the support of $O_A(t)$ is $\Lambda$. Define $S=\{j\in\Lambda\mid \operatorname{dist}(A,j)\leq L^\ast/4\}$ , and $\bar{S}=\Lambda-S$ is the complement of $S$. Following Ref. \cite{liebrobinson2}, we can approximates $O_A(t)$ by an operator $O_{A}^{L^\ast/4}(t)=\frac{1}{\operatorname{Tr}_{\bar{S}}\left(\mathbbm{1}_{\bar{S}}\right)} \operatorname{Tr}_{\bar{S}}\left(O_{A}(t)\right) \otimes \mathbbm{1}_{\bar{S}}$, whose support is $S$. Indeed,
\begin{eqnarray}\label{haarint}
O_{A}^{L^\ast/4}(t)=\int d\mu(V)VO_A(t)V^\dagger,
\end{eqnarray}
where $V$ is a unitary operator acting on $\bar{S}$ and $\mu(V)$ is the Haar measure for $V$. Therefore, $\|O_A(t)-O_A^{L^\ast/4}(t)\|\leq\int d\mu(V)\|[V,O_A(t)]\|$. Combining Eq. (\ref{boundedLC}), and absorbing the geometric factor into $C$,  we have $\|O_A(t)-O_A^{L^\ast/4}(t)\|\leq CL^{\ast D}e^{-\frac{1}{4}\mu L^\ast}$.
Since $\operatorname{diam}(S)\leq L^\ast$ \cite{SMlemma} and $P_0$ has TO to accuracy $(L^\ast,\epsilon)$, there exists a scalar $z$ such that $\|P_0O_A^{L^\ast/4}(t)P_0-zP_0\|\leq\epsilon$.
Applying triangle inequality, we finally get
\begin{eqnarray}
\|P_0O_A(t)P_0-zP_0\|\leq\epsilon+CL^{\ast D}e^{-\frac{1}{4}\mu L^\ast}=\epsilon^\prime.
\end{eqnarray}
Since exponential decay overwhelms algebraic increase, $\epsilon^\prime=\operatorname{exp}(-\Omega(L))$. So $P_0(t)$ has TO to accuracy $(\frac{L^\ast}{2},\epsilon^\prime)$.
\end{proof}}

\begin{proposition}\label{P2}
For a local Hamiltonian $H_0=\sum_{Z}H_Z$ where $Z$'s are bounded sets, $H(t)=U(t)H_0U(t)^\dagger$ defines a family of iso-spectral local Hamiltonians.
\end{proposition}
\begin{proof}
Since $U(t)$ is unitary, $H(t)$ is iso-spectral for all $t$. We next prove $H(t)=\sum_Z H_Z(t)$ is a local Hamiltonain.
First, as in Eq. (\ref{haarint}), each $H_Z(t)$ can be approximated by $H^l_Z(t)=\int\!\!d\mu(V)\;VH_Z(t)V^\dagger$,
where $V$ is a unitary operator acting on the set with distance larger than $l$ from set $Z$. Denote by $B_l(Z)$ the support of $H^l_Z(t)$. 
By Eq. (\ref{boundedLC}), we get
\begin{eqnarray}\label{localapproxiamtion}
\left\|H_{Z}(t)-H_{Z}^{l}(t)\right\| \leq C\left|Z\right|\left\|H_Z\right\|e^{-\mu l}.
\end{eqnarray}
In general, $H=\sum_{Z^\prime} H_{Z^\prime}$ is a local Hamiltonian if for any point $j\in\Lambda$,
\begin{eqnarray}\label{localdefinition}
\sum_{Z^\prime \ni j}\left\|H_{Z^\prime}\right\||Z^\prime| \exp [\nu \operatorname{diam}(Z^\prime)]=\mathcal{O}(1),
\end{eqnarray}
where $\nu$ is a positive constant \cite{Hastings2006,Bravyi2010a,ChenXie2010}. Here $\operatorname{diam}(Z^\prime)$ can be arbitrary large, while $\left\|H_{Z^\prime}\right\|$ needs to be exponentially decaying with $\operatorname{diam}(Z^\prime)$. This general notion of locality allows the interaction term can have an exponentially decaying tail instead of being exactly finite support.
We decompose $H_Z(t)=\sum_l\tilde{H}_Z^l(t)$ by defining a sequence of operators $\tilde{H}_Z^l(t)=H_Z^{l}(t)-H_Z^{l-1}(t),~\tilde{H}_Z^0=H_Z^0(t)$.
$\tilde{H}_Z^l(t)$ is supported on set $B_l(Z)$ with $\operatorname{diam}\left(B_l(Z)\right)\leq\operatorname{diam}(Z)+2l$ \cite{SMlemma}, and its norm can be bounded using Eq. (\ref{localapproxiamtion}) and the triangle inequality: $\left\|\tilde{H}_Z^l(t)\right\|\!\leq\! C^{\prime}\!e^{\frac{\mu}{2}\operatorname{diam}(Z)}\!\left|Z\right|\!\|H_Z\|e^{-\frac{\mu}{2} \operatorname{diam}(\!B_l(Z)\!)}\!$,
where $C^\prime=C(1+e^{\mu})$ is a constant. Since $\left|Z\right|$, $\|H_Z\|$ and $\operatorname{diam}(Z)$ are bounded by constants, $H(t)=\sum_Z H_Z(t)=\sum_{Z,l}\tilde{H}_Z^l(t)$, satisfying local condition Eq. (\ref{localdefinition}), is a local Hamiltonian.
\end{proof}

Proposition \ref{P1} states that the initial ground space continues to be a QEC with macroscopic distance after time evolution if the system is of a BLC. Proposition \ref{P2} further manifests that all $H(t)$ belong to the same connected component of iso-spectral local Hamiltonians so that adiabatic connection is well defined \cite{adiabatically,ChenXie2010}, so the initial quantum phase of TO is preserved \cite{ChenXie2010}. As concrete examples, we will show how the BLCs can emerge in perturbed TO models.

{\em Kitaev's toric code and honeycomb model with BLCs.} Consider first the two-dimensional toric code \cite{kitaev:2003} defined on an $L\times L$  square lattice $\Lambda$ with periodic boundary conditions, where spins $1/2$ reside on the bonds (equivalently at sites in Ref. \cite{Wen2003}) of the lattice.
The Hamiltonian is the sum of mutual-commuting four-body interactions: $H_{TC}(J)=-\sum_{s}J_{s}A_{s}-\sum_{p}J_{p}B_{p}$,
where $A_{s}\equiv\prod_{i\ni s}\sigma_{i}^{x}$ and { $B_{p}\equiv\prod_{i\in\partial p}\sigma_{i}^{z}$} are stabilizer operators indexed  by $s$ on the lattice site (vertex) and $p$ on the dual lattice site (face). All the coupling constants $J_{s}$ and $J_{p}$ are positive, so each stabilizer operator acts as $+1$ on the 4-fold degenerate ground space.
\begin{figure}
\centering
\includegraphics[width=0.4\textwidth]{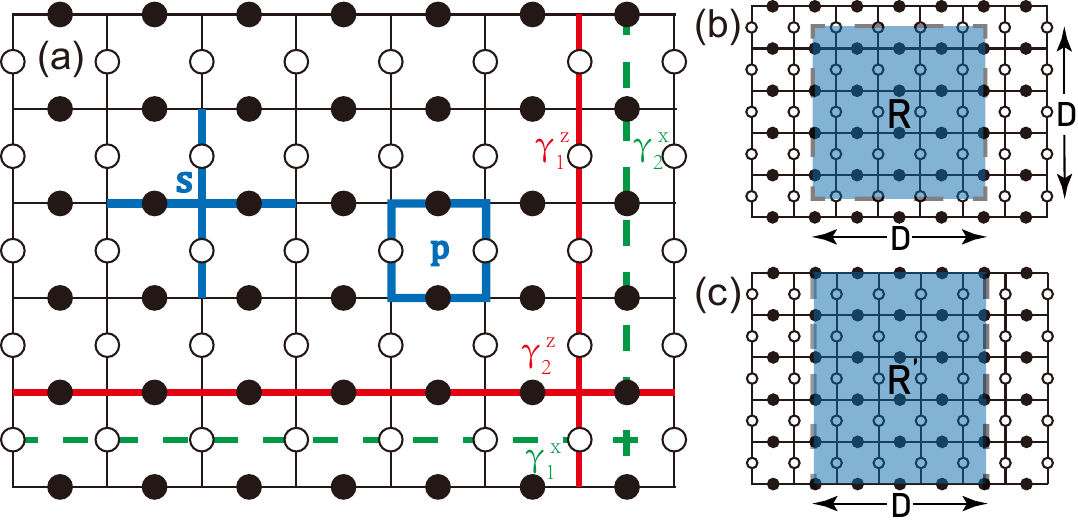}\\
\caption{ (a) Illustration of the square lattice $\Lambda$ with physical spins living on the bonds in odd rows (black dots) and even rows (white dots). The Examples of star (s), plaquete (p), and the non-contractible path $\gamma^\alpha_i$ ($\alpha=x,z$ and $i=1,2$) are shown. (b) Square $D\times D$ region $R$. (c) Cylindrical $L\times D$ region $R^\prime$ }\label{lattice}
\end{figure}

The protocol of dynamics we consider is quantum quench \cite{Mitra2018}. The initial state is $|\Psi(0)\rangle$, a ground state of pre-quench Hamiltonian $H_{TC}$ (without loss of generality, we choose the sector of $W_1^x=1, W_2^z=1$ \cite{note1}), and the post-quench Hamiltonian reads
\begin{eqnarray}\label{Hsigma}
\!\!\!\!\!H(J,h)\!\!=H_{TC}(J)-\!\!\!\sum_{\substack{i\in \text{odd}\\\text{rows}}}\!\!h^{o}_{i}\sigma^{z}_{i}\!-\!\!\!\sum_{\substack{j\in \text{even}\\\text{rows}}}\!\!h^{e}_{j}\sigma^{x}_{j},
\end{eqnarray}
where the odd (even) rows are shown in Fig. \ref{lattice} (a).  Then the initial state evolves as $|\Psi(t)\rangle=U(J,h;t)|\Psi(0)\rangle$, with
$U(J,h;t)=e^{-itH(J,h)}$.
We can map the stabilizer operators to effective spins residing on lattice and dual lattice sites \cite{Yu2016,toric2Ising}: $A_s\mapsto\tau^z_s$ and $B_p\mapsto\tau^z_p$. 
In this `$\tau$-picture', Eq. (\ref{Hsigma}) is the sums of independent quantum Ising chains: $H(J,h)=\sum_{l=1}^{2L}\sum_{j=1}^{L}\left[-J_{l,j}\tau^z_{l,j}-h_{l,j}\tau^x_{l,j}\tau^x_{l,j+1}\right]$,
with period boundary condition in the sector we choose. $H(J,h)$ can be solved via Jordan-Wigner transformations: $\tau^z_j=1-2c_j^\dagger c_j$ and $\tau^x_j=\prod_{i<j}(1-2c_i^\dagger c_i)(c_j+c_j^\dagger)$, where we have omitted the row index. The Hamiltonian in each row is quadratic: $H_l(J,h)=\frac{1}{2}\psi^\dagger\mathcal{H}(J,h) \psi$, where $\psi^\dagger=\left(c_1^\dagger, c_1, c_2^\dagger, c_2,\cdots,c_L^\dagger, c_L \right)$. The first quantized Hamiltonian is given as a $2\times2$-block tridiagonal Jacobi matrix (except for the boundary terms) $\mathcal{H}(J,h)_{m,n}=(2J_m\delta_{m,n}-h_m\delta_{m,n-1}-h_n\delta_{m-1,n})\sigma^z-(h_{m}\delta_{m,n-1}-h_{n}\delta_{m-1,n})i\sigma^y$, and the boundary condition is antiperiodic for the sector we choose.
In the Heisenberg picture, $c_i(t)\!=\!\!\sum_{j=1}^{N}\mathcal{U}_{2i-1,2j-1}(t)c_{j}\!+\!\mathcal{U}_{2i-1,2j}(t)c_{j}^\dagger$, where $\mathcal{U}(t)=e^{-it\mathcal{H}}$.
Refs. \cite{Hamza2012, mathlocalizationXY2} proves that the system is of BLC provided the \emph{dynamical-localization} condition is satisfied: 
\begin{eqnarray}\label{oneparticlebound}
\!\mathbb{E}\left[\!\sup _{t\in\mathbb{R}}\left(\mathcal{M}_{n,m}(t)\right)\!\right]\!\leq\! C\!e^{-\mu\operatorname{dist}\left(n, m\right)}\!.
\end{eqnarray}
Here, $\mathcal{M}_{n,m}(t)=\left|\mathcal{U}_{2n\!-\!1,2m\!-1\!}(t)\right|\!+\!\left|\mathcal{U}_{2n\!-\!1,2m\!}(t)\right|$ \cite{typicalvalue}. The exponential decay in Eq. (\ref{oneparticlebound}) is proved in Ref. \cite{mathlocalizationXY1} at large disorder and sufficiently smooth distribution of $\{J\}$. For arbitrary nontrivial compactly supported distributions, Ref. \cite{mathlocalizationXY2} prove that the bound decays sub-exponentially provided the gap is not closed.

\begin{figure}
\center
\includegraphics[width=0.9\linewidth]{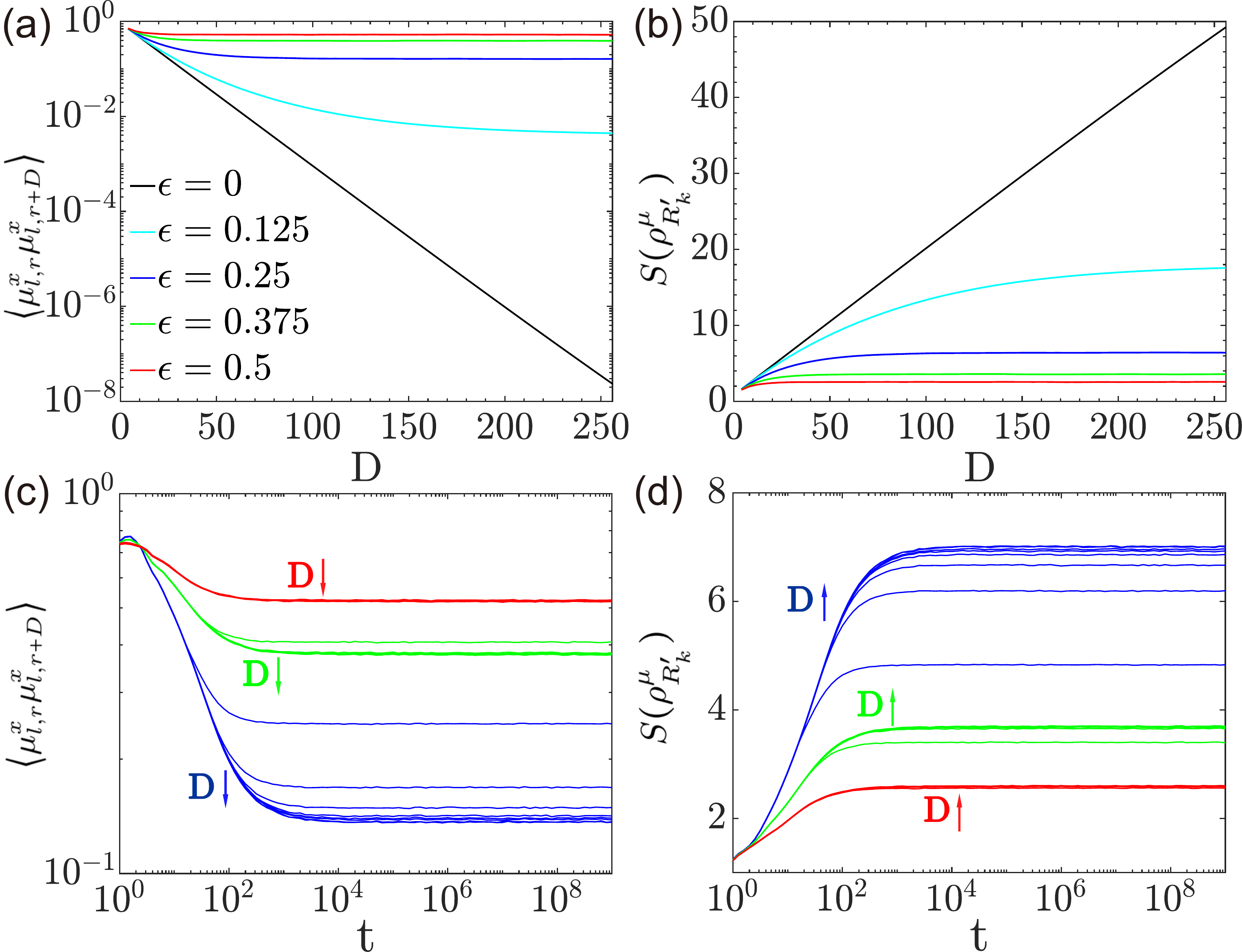}\\
\caption{ Disorder-averaged (a) $\left\langle\mu_{l,r}^{x} \mu_{l,r+D}^{x}\right\rangle$; (b) $S(\rho^\mu_{R^\prime_k})$ at $t=250$ for each $\epsilon$ with 1000 disorder realization, $L=1024$. Disorder-average for each $\epsilon$ with 2000 disorder realization of (c) $\left\langle\mu_{l,r}^{x} \mu_{l,r+D}^{x}\right\rangle$ with increasing $D$ from top to bottom; (d) $S(\rho^\mu_{R^\prime_k})$ with increasing $D$ from bottom to top after long time evolution, $D\in\{32,64,\cdots,256\}$, $L=512$.}\label{WL-EE}
\end{figure}

Setting $J_j=1+\epsilon \eta_j$ where $\eta_j\in[-1,1]$ are i.i.d random variables and $h_j=0.5$, we investigate two typical \emph{nonlocal} order parameters for TO to confirm $|\Psi(t)\rangle$ and $|\Psi(0)\rangle$ belong to the same phase. For $h^o\neq0$ and $h^e=0$, the $Z_2$ gauge structure is intact during the time evolution. For a $D\times D$ square region $R$ with boundary $\partial R$, the \emph{Wilson loop} operator reads ${W}_{\partial R} \equiv \prod_{i \in \partial R} \sigma_{i}^{x}=\prod_{s \in R} A_{s}=\prod_{s \in R} \tau^z_{s}$ \cite{Halasz2012}, see Fig. \ref{lattice} (b).
Introducing $\mu_{l,j}^{x}=\prod_{k \leq j}\tau_{l,k}^{z}$, we have $\left\langle{W}_{\partial R}\right\rangle=\prod^D_{l=1}\left\langle\mu_{l,r}^{x} \mu_{l,r+D}^{x}\right\rangle$ in this $\mu$ picture. The numerical results shown in Fig. \ref{WL-EE} (a, c) indicate that, as disorder increases, $\left\langle\mu_{l,r}^{x} \mu_{l,r+D}^{x}\right\rangle$ tends to show greater resilience; over a long-time scale, it converges as $D$ increases. These result in the perimeter law: $\left\langle{W}_{\partial R}\right\rangle\sim\operatorname{exp}\left(-\mathcal{O}(|\partial R|)\right)$, where $|\partial R|=4D$ is the cardinality of the set of the boundary. This implies a deconfined (TO) phase.


The \emph{topological entanglement entropy} \cite{hamma:2005b,hiz1,hamma2005,kitaevpreskill,levin:2006} is another nonlocal order parameter for TO. The entanglement entropy of the reduced density operator in the $\sigma$ picture, $\rho_{R^\prime}^{\sigma}$, equals the sum of entanglement entropy of each row in the $\mu$ picture \cite{SM}: $S(\rho_{R^\prime}^{\sigma})=\sum_{k=1}^{2L}S(\rho_{R^\prime_k}^{\mu})$, where $R^\prime$ is a cylindrical subsystem shown in Fig. \ref{lattice} (c). The boundary of $R^\prime$ contains only left and right sides at a distance of $D$, and the length of each side is $L$.
For the ground state $\rho_0$ of the TCM, $S(\rho_{0R}^{\sigma})=2L=|\partial R^\prime|$ \cite{hamma:2005b,hiz1,hamma2005}, yet the topological entropy seems absent. This paradox arises from the selection of ground state which is an equal-weighted superposition of all contractible loops and a non-contractible loop crossing the boundary. However, the ground state $\rho^\prime_0$ in sector $W_1^z=1, W_2^z=1$ contains only contractible loops with $S(\rho_{0R^\prime}^{\prime\sigma})=|\partial R^\prime|-1$, where the topological entropy appears as $\operatorname{log}_{2}2=1$.
The numerical results shown in Fig. \ref{WL-EE} (b, d) manifest the entanglement boundary law:  $S(\rho_{R^\prime}^{\sigma};\epsilon)=\sum_{k=1}^{2L}S(\rho_{R^\prime_k}^{\mu};\epsilon)\leq 2\alpha(\epsilon)L$, where $\alpha(\epsilon)$ converges as $D$ increases. For the same reason as the ground state, the topological entanglement entropy is $\operatorname{log}_{2}2=1$ in the thermodynamic limit for both system and subsystem.

\begin{figure}
\center
\includegraphics[width=0.9\linewidth]{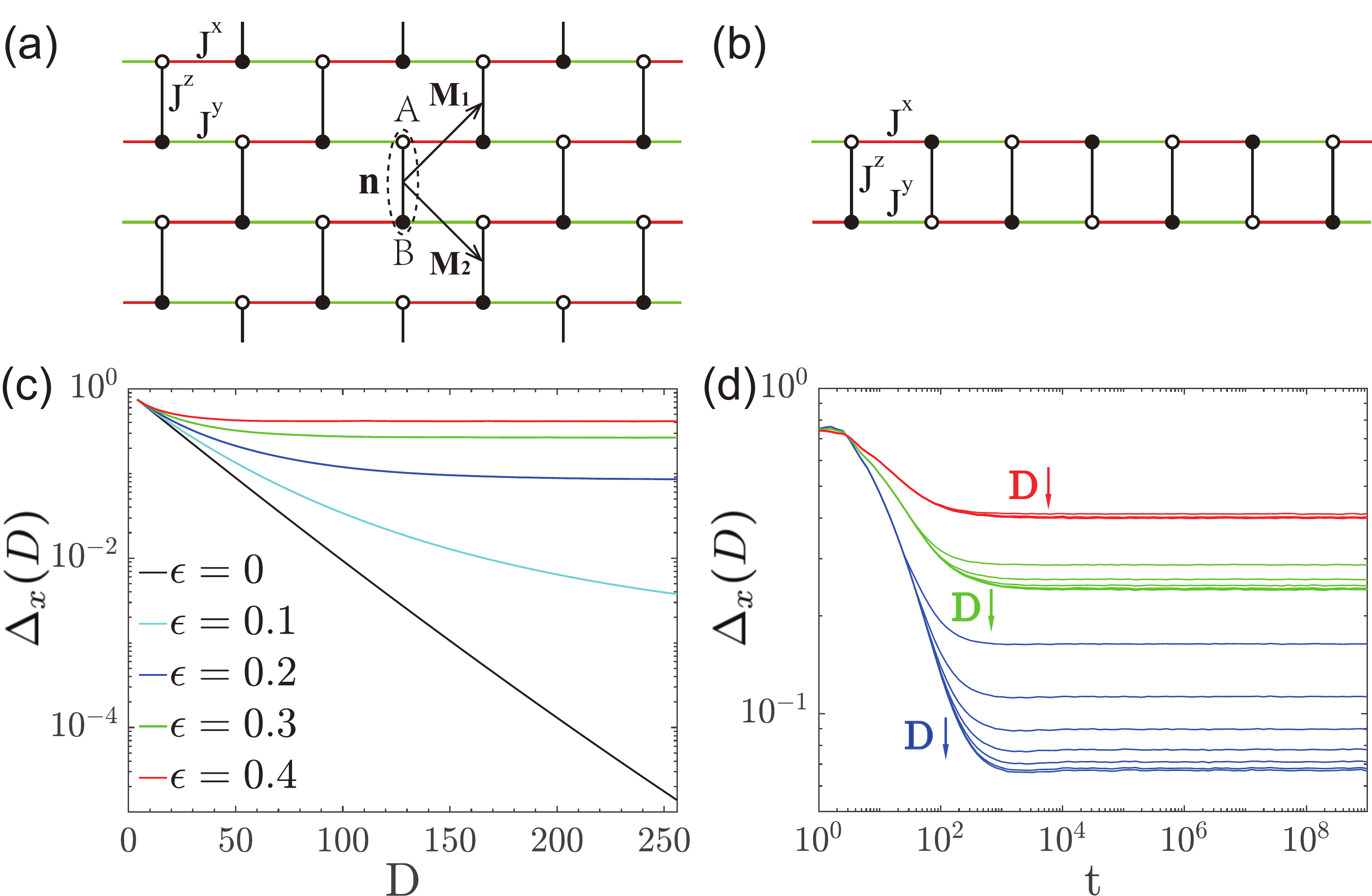}\\
\caption{(a) Illustration of Kitaev's honeycomb model defined on a brick-wall lattice with spins living at sites. The unit cell $\mathbf{n}$ contains two sites belonging to sublattices $A$ (white dot) and $B$ (black dot), respectively. Three types of couplings are denoted in red (x), green (y), and black (z). (b) The ladder model reduced from the honeycomb model. (c) Disorder-averaged $\Delta_x(D)$ at $t=250$ with $1000$ disorder realizations for each $\epsilon$, $L=1024$. (d) Disorder-averaged $\Delta_x(D)$ with increasing $D$ from top to bottom after long time evolution, $D\in\{64, 96,\cdots,256\}$; $2000$ disorder realizations for each $\epsilon$, $L=512$.}\label{honeycomb_ladder_SOP}
\end{figure}

In the end, we discuss Kitaev's honeycomb model \cite{Kitaev2006}, where spins reside on the sites of a honeycomb (brick-wall) lattice with periodic boundary conditions shown in Fig. \ref{honeycomb_ladder_SOP} (a). The honeycomb model is an approximate QEC in the same phase as the toric code \cite{Lee2017}, while the Hamiltonian comprises only (non-commuting) \emph{two-body interactions}: $H_{\!K\!H\!}=-\sum_{\mathbf{n}}J_{\mathbf{n}}^x\sigma_{(\mathbf{n},A)}^{x}\sigma_{(\mathbf{n+M_1},B)}^{x}+
J_{\mathbf{n}}^y\sigma_{(\mathbf{n},B)}^{y}\sigma_{(\mathbf{n+M_2},A)}^{y}+J_{\mathbf{n}}^z\sigma_{(\mathbf{n},A)}^{z}\sigma_{(\mathbf{n},B)}^{z}$. Here, $\mathbf{n}$ is the position vector of the unit cell; $\mathbf{M_1}$ and $\mathbf{M_2}$ are primitive translation vectors; $A$ ($B$) represents the sublattice. The honeycomb model can be solved via Jordan-Wigner transformations \cite{Feng2007, ChenHu2007}: $\sigma_{l,j}^z=1-2c^\dagger_{l,j}c_{l,j}$ and $\sigma_{l,j}^x=\prod_{m<l}\prod_{k}(1-2c^\dagger_{m,k}c_{m,k})\prod_{k<j}(1-2c^\dagger_{l,k}c_{l,k})(c_{l,j}+c^\dagger_{l,j})$. The fermion representation of the Hamiltonian is $H_{\!K\!H\!}=i\sum_{\mathbf{n}}J_{\mathbf{n}}^x d_{(\mathbf{n},A)}d_{(\mathbf{n+M_1},B)}-J_{\mathbf{n}}^y d_{(\mathbf{n},B)}d_{(\mathbf{n+M_2},A)}-G_{\mathbf{n}}J_{\mathbf{n}}^zd_{(\mathbf{n},A)}d_{(\mathbf{n},B)}$, where $d_{(\mathbf{n},A)}=-i\left(c_{(\mathbf{n},A)}-c^\dagger_{(\mathbf{n},A)}\right)$ and $d_{(\mathbf{n},B)}=c_{(\mathbf{n},B)}+c^\dagger_{(\mathbf{n},B)}$ are Majorana fermion operators. $G_{\mathbf{n}}=\left(c_{(\mathbf{n},A)}+c^\dagger_{(\mathbf{n},A)}\right)\left(c_{(\mathbf{n},B)}-c^\dagger_{(\mathbf{n},B)}\right)$ commutes with the Hamiltonian and acts as a local gauge field. The ground state is in the zero-flux phase \cite{Lieb1994}, so we can set $G_{\mathbf{n}}=1$ for all $\mathbf{n}$. $H_{\!K\!H\!}$ has a quadratic form and its corresponding first quantized Hamiltonian is $\mathcal{H}_{\!K\!H\!}$. It can be diagonalized in the momentum space if translation symmetry is preserved \cite{isingbook}. The model contains three disconnected gapped phase and one gapless phase. Consider TO gapped phase $A_z$ with $|J^z|>|J^x|+|J^y|$ and $J^x, J^y\neq0$ \cite{Kitaev2006,Lee2017}. The BLC emerges when \emph{dynamical-localization} condition is fulfilled:
\begin{eqnarray}\label{oneparticlebound2}
\!\mathbb{E}\left[\!\sup _{t\in\mathbb{R}}\left(\parallel P_{\mathbf{n}}\operatorname{e}^{-it\mathcal{H}_{KH}}P_{\mathbf{m}}\parallel\right)\!\right]\!\leq\! C^\prime\!e^{-\mu^\prime\operatorname{dist}\left(\mathbf{n}, \mathbf{m}\right)}\!,
\end{eqnarray}
where $P_{\mathbf{n}}$ is the projector onto the Hilbert space at site $\mathbf{n}$ and $\parallel\cdot\parallel$ is the norm of a $2\times2$ matrix. A general result of Ref. \cite{mathlocalizationXY1} covers the model we discussed, from which Eq. (\ref{oneparticlebound2}) is valid when $\{J^z_{\mathbf{n}}\}$ are at large disorder with sufficiently smooth distribution.

By reducing the honeycomb model to a ladder model [Fig. \ref{honeycomb_ladder_SOP} (b)] \cite{Feng2007,Hastings2021}, the vector $\mathbf{n}$ degenerates to a number $n$ and the lattice vectors $\mathbf{M}_1=\mathbf{M}_2$ degenerate to $1$ . The fermion representation of the ladder-model Hamiltonian is $H_{LM}=i\sum_{n=1}^{L}J_{n}^x d_{(n,A)}d_{(n+1,B)}-J_{n}^y d_{(n,B)}d_{(n+1,A)}-G_{n}J_{n}^zd_{(n,A)}d_{(n,B)}$. We set $G_{n}=(-1)^n$ since the ground state is in the $\pi$-flux phase \cite{Lieb1994}. The ladder model in the original spin space has no local order parameter, but can be characterized by nonlocal string order parameters (SOPs) \cite{Feng2007}. For gapped phase $A_x$, The SOP is $\Delta_x(D)=|\langle\prod_{n=1}^{D}\sigma^x_{n,A}\sigma^x_{n+1,B}\rangle|=|\langle\prod_{n=1}^{D}id_{n,A}d_{n+1,B}\rangle|$, and TO asks $\operatorname{lim}_{D\rightarrow\infty}\Delta_x(D)>0$. Set the pre-quench Hamiltonian with parameters $\{J_n^x=1, J_n^y=0.1, J_n^z=0.1\}$ and the post-quench Hamiltonian $\{J_n^x=1+\epsilon\eta_n, J_n^y=0.1, J_n^z=0.5\}$, where $\eta_n\in[-1, 1]$ are i.i.d random variables. 
Keeping $t=250$ fixed, the numerical results shown in Fig. \ref{honeycomb_ladder_SOP} (c) indicate that $\Delta_x(D)$ tends to zero exponentially with increasing $D$ when $\epsilon=0$; as the disorder increases, $\Delta_x(D)$ shows increasing resilience. For long-time evolution, Fig. \ref{honeycomb_ladder_SOP} (d) shows that $\Delta_x(D)$ converges to a positive value as $D$ increases. The results imply that the initial TO is preserved after a quantum quench by introducing suitable random coupling \cite{LadderDL}.


{\em Discussion.} The limitations of the models should be discussed. The special random perturbations render the Hamiltonians quadratic to satisfy the dynamical-localization condition, while other random perturbations may induce interaction terms \cite{Zhu2021,Venn2022}. Numerical evidence \cite{Venn2022} shows topological local integrals of motion \cite{Wahl2020} emerge in the toric code with small random unidirectional fields that preserve $Z_2$ gauge, leading to a logarithmic light cone \cite{Kim2014} (the lifetime of TO is exponentially long \cite{Zeng2023}). However, this will not be the case for the non-Abelian topological phase of the honeycomb model according to the argument in \cite{Potter2016}. 

Self-correcting QMs must be robust against thermal noise and small imperfections in the ideal Hamiltonian \cite{rmp2016}. For the first type of noise, except in four or higher spatial dimensions \cite{dennis,alicki,hammamazac,Shen2022}, large classes of topological-QM models are unfortunately not self-correcting \cite{nogotheorem2d1,nogotheorem2d2,nogotheorem3d,membrane,finiteT,chamon3d,nussinov,chesitherm,iblisdir,Hastings2011,hammamazac,Mohseninia2016},
though polynomial lifetimes for the memories can be obtained through sophisticated designs \cite{toricboson,chesi-mem,haah2011,bravyihaah2013}. For the second type, quantum dynamics render the topological QMs unreliable even without a thermal bath \cite{Kay2009,Pastawski2010,kay2011,Bravyi2012}; however, we demonstrate that TOs and QMs can be dynamically robust at zero temperature with BLCs, in which regime the time evolution is a quasi-adiabatic continuation. This potentially offers a new perspective on the exploration of topological QMs simultaneously suffering both types of noise. For instance, under which conditions can BLCs (or logarithmic light cones) emerge for an open system \cite{Poulin2010,Nachtergaele2011,Breteaux2023,Trivedi2024}?


\begin{acknowledgments}
This work was supported by National Key R\&D Program of China (grant Nos. 2021YFA1400900, 2021YFA0718300, 2021YFA1400243), NSFC (grant No. 61835013) (W. -M. L.); the National Key R\&D Program of China (grant No. 2017YFA0304300), Strategic Priority Research Program of the Chinese Academy of Sciences (grant No. XDB28000000)(H. F.); NSFC (Grant Nos. 12074410, 12047502, 11934015)(J. -P. C); the JSPS Postdoctoral Fellowship (Grant No.~P19326), the JSPS KAKENHI (Grant No.~JP19F19326)(Y. -R. Z); NSF (award No. 2014000) (A. H.).
\end{acknowledgments}


%

\end{document}


\title{Supplemental Material:\\Bounded light cone and robust topological order out of equilibrium}
\author{Yu~Zeng}\thanks{zengyunk@gmail.com}
\affiliation{Beijing National Laboratory for Condensed Matter Physics, Institute of Physics, Chinese Academy of Sciences, Beijing 100190, China}
\affiliation{Shandong Inspur Intelligence Research Institute Co., Ltd, Jinan, China}
\author{Alioscia~Hamma}\thanks{Alioscia.Hamma@umb.edu}
\affiliation{Department of Physics, University of Massachusetts Boston, 100 Morrissey Blvd, Boston MA 02125, USA}
\affiliation{Dipartimento di Fisica `Ettore Pancini', Universit\`a degli Studi di Napoli Federico II, Via Cintia 80126, Napoli, Italy}
\author{Yu-Ran Zhang}
\affiliation{Theoretical Quantum Physics Laboratory, RIKEN Cluster for Pioneering Research, Wako-shi, Saitama 351-0198, Japan}
\affiliation{School of Physics and Optoelectronics, South China University of Technology, Guangzhou 510640, China}
\author{Jun-Peng Cao}
\affiliation{Beijing National Laboratory for Condensed Matter Physics, Institute of Physics, Chinese Academy of Sciences, Beijing 100190, China}
\author{Heng~Fan}\thanks{hfan@iphy.ac.cn}
\affiliation{Beijing National Laboratory for Condensed Matter Physics, Institute of Physics, Chinese Academy of Sciences, Beijing 100190, China}
\author{Wu-Ming Liu}\thanks{wliu@iphy.ac.cn}
\affiliation{Beijing National Laboratory for Condensed Matter Physics, Institute of Physics, Chinese Academy of Sciences, Beijing 100190, China}

\maketitle



\appendix
\setcounter{table}{1}
\renewcommand{\thetable}{S\arabic{table}}
\appendix
\renewcommand{\theequation}{\thesection.\arabic{equation}}
\appendix
\renewcommand\thefigure{\Alph{section}\arabic{figure}}

\tableofcontents

\section{A lemma applied in the proofs of Proposition 1 and 2}
\begin{mylemma}
A bounded subset $A\subset\Lambda$ has diameter $\operatorname{diam}(A)$. Define a set $S=\{j\in\Lambda\mid \operatorname{dist}(A,j)\leq l\}$, then $\operatorname{diam}(S)\leq\operatorname{diam}(A)+2l$.
\end{mylemma}
\begin{proof}
The diameter of $S$ is defined as $\operatorname{diam}(S)=\operatorname{max}_{i, j\in S}\operatorname{dist}\left(i, j\right)$. So there exist two sites $a_1$ and $b_1$ in $S$ such that $\operatorname{dist}(a_1, b_1)=\operatorname{diam}(S)$. For $a_1$, we can find $a_2\in A$ such that $\operatorname{dist}(a_1, a_2)=\operatorname{min}_{j\in A}\operatorname{dist}(j, a_1)$. Considering the definition of $S$ is $S=\{j\in\Lambda\mid \operatorname{dist}(A,j)\leq l\}$, we have $\operatorname{dist}(a_1, a_2)\leq l$. Similarly, we can find $b_2\in A$ corresponding to $b_1$ such that $\operatorname{dist}(b_1, b_2)=\operatorname{min}_{j\in A}\operatorname{dist}(j, b_1)$, and we have $\operatorname{dist}(b_1, b_2)\leq l$. We can apply triangle inequality several times:
\begin{eqnarray}
\operatorname{diam}(S)&=&\operatorname{dist}(a_1, b_1)\nonumber\\
&\leq&\operatorname{dist}(a_1, a_2)+\operatorname{dist}(a_2, b_1)\nonumber\\
&\leq&\operatorname{dist}(a_1, a_2)+\operatorname{dist}(b_2, b_1)+\operatorname{dist}(a_2, b_2)\nonumber\\
&\leq& 2l+\operatorname{diam}(A)\nonumber
\end{eqnarray}
\end{proof}

\section{Numerical examples of $\mathcal{M}_{n,m}(t)$ in the main text}
\begin{figure}
\centering
\includegraphics[width=\linewidth]{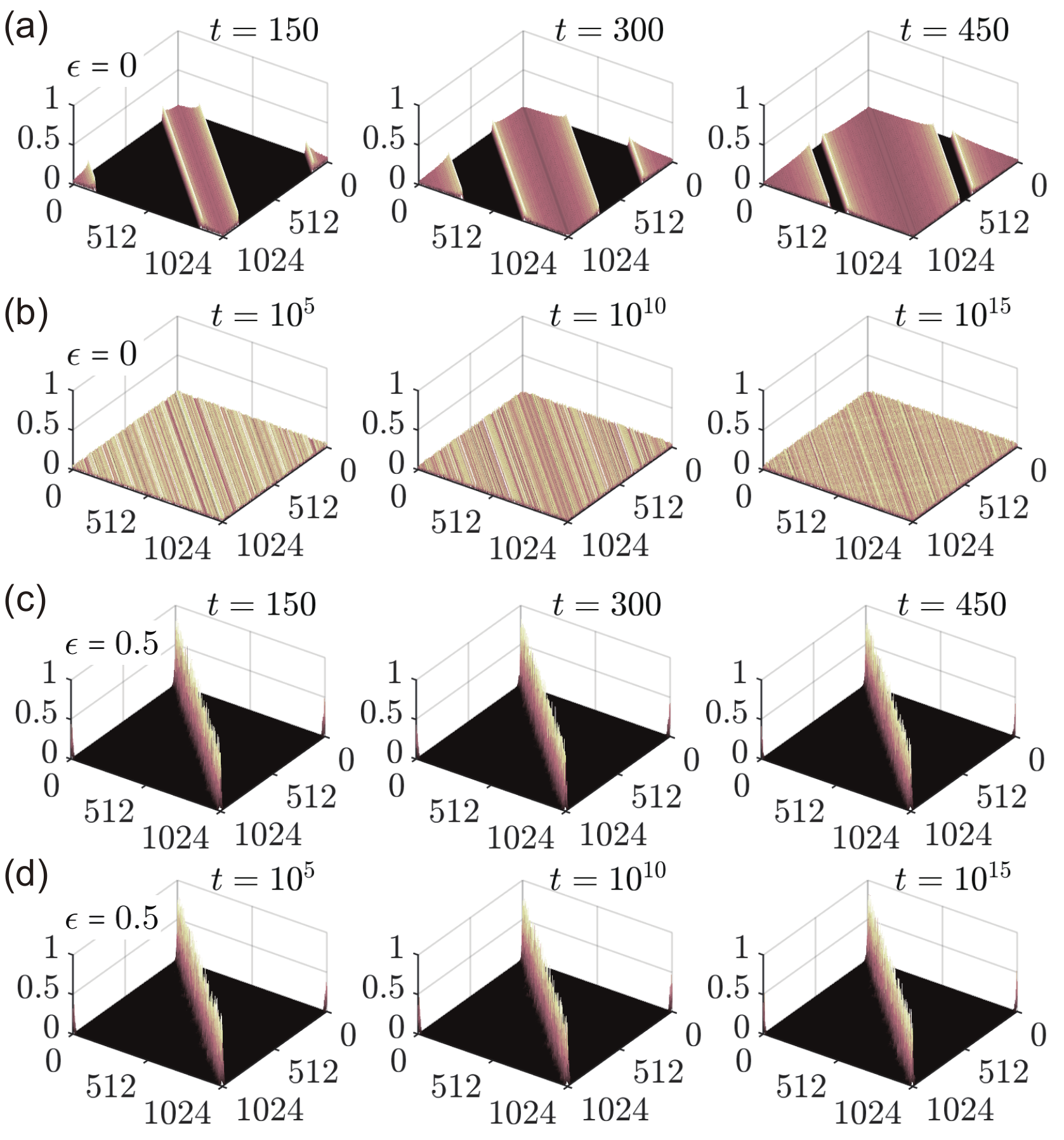}\\
\caption{ Typical values of $\mathcal{M}_{n,m}(t)=\left|\mathcal{U}_{2n\!-\!1,2m\!-1\!}(t)\right|\!+\!\left|\mathcal{U}_{2n\!-\!1,2m\!}(t)\right|$ for (a)-(b) $\epsilon=0$ and (c)-(d) $\epsilon=0.5$.}\label{Umatrix}
\end{figure}

We present typical numeric values of $\mathcal{M}_{n,m}(t)$ appearing in Eq. (8) in the main text. Here, $\mathcal{M}_{n,m}(t)=\left|\mathcal{U}_{2n\!-\!1,2m\!-1\!}(t)\right|\!+\!\left|\mathcal{U}_{2n\!-\!1,2m\!}(t)\right|$ where $\mathcal{U}(t)=e^{-it\mathcal{H}}$. Setting $J_j=1+\epsilon \eta_j$ where $\eta_j\in[-1,1]$ are i.i.d random variables, and $h_j=0.5$, we illustrate the numerical results of $\mathcal{M}_{n,m}(t)$ with $\epsilon=0$ and $\epsilon=0.5$ for different time scales in Fig. \ref{Umatrix}. For the clean case $\epsilon=0$, the peaks of $\mathcal{M}(t)$ spread in the anti-diagonal direction with a linear velocity and are uniformly distributed in every matrix entry on long-time scales. So every local operator in Heisenberg picture will be nonlocal after a long-time evolution. For example, $H(t)$ in proposition {\bf 2} in the main text will not be a local Hamiltonian. In contrast, disorders with nonzero $\epsilon$ make $\mathcal{M}(t)$ close to the identity matrix with exponentially small off-diagonal elements at all times. As a result, the local operator will always be quasi-local with exponentially small tail.
\section{Von Neumann entanglement entropy for a cylindrical region}
For the $L\times L$ square lattice with periodic boundary condition, we consider the entanglement entropy between a cylindrical subsystem ${R^{\prime}}$ and its complement, see Fig. 2(c) in the main text. For arbitrary density matrix $\rho$, the reduced density operator can be expressed as \cite{Latorre2003,Jin2004}
\begin{eqnarray}\label{reduceddensitymatrix}
\rho_{{R^{\prime}}}=2^{-L(2D+1)}\!\!\!\!\!\!\!\!\!\!\!\!\sum_{\substack{\alpha_j\in\{0,x,y,z\}\\j\in R^\prime}}\prod_{j\in{R^{\prime}}}\sigma^{\alpha_j}_j\operatorname{tr}[(\prod_{j\in{R^{\prime}}}\sigma^{\alpha_j}_j)^\dagger\rho].
\end{eqnarray}
The normalization coefficient $2^{-N(2D+1)}$ results from the dimension of spin space and the number of spins in the ${R^{\prime}}$. For an arbitrary pure state $\rho=|\Psi\rangle\langle\Psi|$, Eq. (\ref{reduceddensitymatrix}) is
\begin{eqnarray}\label{reduceddensitymatrix2}
\rho_{{R^{\prime}}}=2^{-L(2D+1)}\!\!\!\!\!\!\!\!\!\!\!\!\sum_{\substack{\alpha_j\in\{0,x,y,z\}\\j\in R^\prime}}\prod_{j\in{R^{\prime}}}\sigma^{\alpha_j}_j\langle\Psi|(\prod_{j\in{R^{\prime}}}\!\!\sigma^{\alpha_j}_j\!)^\dagger|\Psi\rangle.
\end{eqnarray}
As mentioned in the main text, $W_{1}^{x} =\prod_{l\in \gamma_{1}^{x}}\sigma_l^x$ and $W_{2}^{z} =\prod_{l\in \gamma_{2}^{z}}\sigma_l^z$ commute with the Hamiltonian, and $|\Psi\rangle$ is in the sector of  $W_1^x=1$, $W_2^z=1$. Here, the non-contractible path $\gamma^x_1$ ($\gamma^z_2$) can be arbitrary even (odd) closed horizontal line. We note that if an operator $O$  anti-commutes with any  $W_1^x$ or $W_2^z$, $\langle\Psi|O|\Psi\rangle=0$. For this reason, the reduced density operator can be written as
\begin{eqnarray}\label{reduceddensitymatrix3}
\rho_{{R^{\prime}}}=2^{-L(2D+1)}\!\!\!\!\!\!\!\!\!\!\!\!\sum_{\substack{g\in G_{{R^{\prime}}}, h\in H_{{R^{\prime}}}\\ x\in X_{{R^{\prime}}}, z\in Z_{{R^{\prime}}}}}\!\!\!\!\!\langle\Psi|xhzg|\Psi\rangle g_{{R^{\prime}}}z_{{R^{\prime}}}h_{{R^{\prime}}}x_{{R^{\prime}}},
\end{eqnarray}
where the notation follows Ref. \cite{Yu2016}. We first define 4 groups denoted by $G$, $H$, $X$ and $Z$. $G$ is generated by all independent $A_s=\prod_{j\ni s}\sigma^x_j$; $H$ is generated by all independent $B_p=\prod_{j\in p}\sigma^z_j$; $X$ is generated by all $\sigma^x$ on the bonds belonging to even rows; and $Z$ is generated by all $\sigma^z$ on the bonds belonging to odd rows.  Then the subgroup $G_{{R^{\prime}}}$ can be defined as $G_{{R^{\prime}}}=\{g\in G|g=g_{R^\prime}\otimes\mathbbm{1}_{\bar{R^\prime}}\}$, where $R^\prime$ denotes the complement of $R$. It means that the elements of $G_{R^\prime}$ are supported on $R^\prime$. The subgroups of $H_{R^\prime}$, $X_{R^\prime}$ and $Z_{R^\prime}$ can be defined in a similar way. The elements of all these groups can be mapped to the $\mu$ picture unambiguously.

The Hamiltonian in the $\mu$ picture,
\begin{eqnarray}\label{Hmu}
H(J,h)=\sum_{l=1}^{2L}\sum_{j=1}^L\left[-J_{l,j}\mu^x_{l,j}\mu^x_{l,j+1}-h_{l,j}\mu^z_{l,j}\right],
\end{eqnarray}
is a sum of uncorrelated quantum Ising chains, so the reduced density operator has the tensor product form
\begin{eqnarray}
\rho^{\mu}_{R^\prime}=\bigotimes_{l=1}^{2L}\rho^{\mu}_{R_l^\prime},
\end{eqnarray}
where $R_l^\prime$ denotes the $l$ th row. Explicitly,
\begin{eqnarray}\label{reduceddensitymatrixodd}
\rho^\mu_{{R_l^{\prime}}}=2^{D+1}\!\!\!\!\!\!\!\!\!\sum_{\substack{g\in G_{{R_l^{\prime}}}, h\in H_{{R_l^{\prime}}}\\ x\in X_{{R_l^{\prime}}}, z\in Z_{{R_l^{\prime}}}}}\!\!\!\langle\Psi^\mu_l|xhzg|\Psi^\mu_l\rangle g_{{R_l^{\prime}}}z_{{R_l^{\prime}}}h_{{R_l^{\prime}}}x_{{R_l^{\prime}}}
\end{eqnarray}
when $l$ is odd; while
\begin{eqnarray}\label{reduceddensitymatrixodd}
\rho^\mu_{{R_l^{\prime}}}=2^{D}\!\!\!\!\!\!\!\!\!\sum_{\substack{g\in G_{{R_l^{\prime}}}, h\in H_{{R_l^{\prime}}}\\ x\in X_{{R_l^{\prime}}}, z\in Z_{{R_l^{\prime}}}}}\!\!\!\langle\Psi^\mu_l|xhzg|\Psi^\mu_l\rangle g_{{R_l^{\prime}}}z_{{R_l^{\prime}}}h_{{R_l^{\prime}}}x_{{R_l^{\prime}}}
\end{eqnarray}
when $l$ is even. As a consequence, the Von Neumann entropy of $\rho^{\sigma}_{R^\prime}$ in the $\sigma$ picture equals the sum of the entropies of $2L$ uncorrelated Ising chains in the $\mu$ piture,
\begin{eqnarray}\label{entanglmententropy}
S(\rho_{R^\prime}^{\sigma})=\sum_{l=1}^{2L}S(\rho_{R^\prime_l}^{\mu}).
\end{eqnarray}

\section{Calculation of correlation function}
The Hamiltonian of each Ising chain in the $\mu$ picture is
\begin{eqnarray}\label{Hmu1d}
\hat{H}(J,h)=\sum_{j=1}^L -J_{j}\mu^x_{j}\mu^x_{j+1}-h_{j}\mu^z_{j}
\end{eqnarray}
with the periodic boundary condition, where the hat is adopted to distinguish from the two dimensional Hamiltonian. We concern the sector of $\prod_{j}\mu_{j}^{z}=1$. The initial state $|\Psi^i_0\rangle$ is a ground of the pre-quench Hamiltonian $\hat{H}^{i}$, and at $t=0$ the Hamiltonian is changed to the post-quench Hamiltonian $\hat{H}^f$, then the initial state will evolve as $|\Psi(t)\rangle=e^{-it\hat{H}^f}|\Psi^i_0\rangle$. We need to calculate the correlation function $\langle\Psi(t)|\mu^x_j\mu^x_l|\Psi(t)\rangle$. We apply the standard method of Jordan-Wigner transformation to map the spins to the free fermions
\begin{eqnarray}\label{Diracfermion}
c_{l}&=&\left(\prod_{j=1}^{l-1}\mu_{j}^{z}\right)\frac{\mu_{l}^{x}+i\mu_{l}^{y}}{2}.
\end{eqnarray}
Then the spin Hamiltonian, Eq. (\ref{Hmu1d}), turns out to be a quadratic fermionic Hamiltonian
\begin{eqnarray}
\label{XYdirac}
\hat{H}(J,h)=&&\sum_{j=1}^L-J_j(c^\dagger_jc_{j+1}-c_jc^\dagger_{j+1}\nonumber\\
&+&c^\dagger_jc^\dagger_{j+1}-c_jc_{j+1})+h_j(c^\dagger_jc_j-c_jc^\dagger_j).
\end{eqnarray}
The general form of a quadratic fermionic Hamiltonian with real parameters is
\begin{eqnarray}
\label{quadratic}
H\!\!=\!\!\frac{1}{2}\!\sum_{m\!n}\!c^\dagger_m\!\mathscr{A}_{m\!n}\!c_n\!-\!c_m\!\mathscr{A}_{m\!n}\!c^\dagger_n\!+\!c^\dagger_m\!\mathscr{B}_{m\!n}\!c^\dagger_n\!-\!c_m\!\mathscr{B}_{m\!n}c_n\!,\nonumber\\
\end{eqnarray}
where $\mathscr{A}_{mn}=\mathscr{A}_{nm}$, and $\mathscr{B}_{mn}=-\mathscr{B}_{nm}$. The Hamiltonian can be diagonalized as
\begin{eqnarray}
H=\frac{1}{2}\sum_k\omega_k(\eta^\dagger_k\eta_k-\eta_k\eta^\dagger_k)=\sum_k\omega_k\eta^\dagger_k\eta_k\!-\!\frac{1}{2}\sum_k\omega_k.
\end{eqnarray}
The quasi-particle operators can be expressed as
\begin{eqnarray}\label{eta2c}
\eta_k=\sum_{l}g_{kl}c_l+h_{kl}c^\dagger_l
\end{eqnarray}
with conditions of
\begin{eqnarray}
\sum_lg_{kl}g_{k^\prime{l}}+h_{kl}h_{k^\prime{l}}&=&\delta_{kk^\prime},\nonumber\\
\sum_lg_{kl}h_{k^\prime{l}}+h_{kl}g_{k^\prime{l}}&=&0.
\end{eqnarray}
Therefore, we can write the diagonalization process as a form of block matrix:
\begin{eqnarray}
\left(
   \begin{array}{cc}
     g & h \\
     h & g \\
   \end{array}
 \right)
 \left(
    \begin{array}{cc}
      \mathscr{A} & \mathscr{B} \\
      -\mathscr{B} & -\mathscr{A} \\
    \end{array}
  \right)
  \left(
   \begin{array}{cc}
     g^T & h^T \\
     h^T & g^T \\
   \end{array}
 \right)=
 \left(
   \begin{array}{cc}
     \omega & 0 \\
     0 & -\omega \\
   \end{array}
 \right)
\end{eqnarray}
and
\begin{eqnarray}
\label{c2eta}
\left(
  \begin{array}{c}
    c \\
    c^\dagger \\
  \end{array}
\right)
=\left(
   \begin{array}{cc}
     g^T & h^T \\
     h^T & g^T \\
   \end{array}
 \right)
 \left(
   \begin{array}{c}
     \eta \\
     \eta^\dagger \\
   \end{array}
 \right).
\end{eqnarray}
Here $\eta$, $\eta^\dagger$, $c$ and $c^\dagger$ are the shorthand notations of columns of fermion operators.

The correlation function is $\langle\Psi(t)|\mu^x_{j}\mu^x_{l}|\Psi(t)\rangle=\langle\Psi(t)| B_{j}A_{j+1}B_{j+1}\cdots A_{l-1}B_{l-1}A_{l}|\Psi(t)\rangle$, where
\begin{eqnarray}\label{AB}
A_{j}=c^\dagger_j+c_j,~~~B_{j}=c^\dagger_j-c_j.
\end{eqnarray}
Here we assume $j<l$ without loss of generality.
Applying the  Wick's theorem, the correlation function above can be expressed as a Pfaffian \cite{barouch:1971}, $|\langle\Psi^i_0|\mu^x_{j}(t)\mu^x_{l}(t)|\Psi^i_0\rangle|=|\operatorname{pf}\Gamma(j,l,t)|$, where the antisymmetric matrix
\begin{eqnarray}
\Gamma(j,l,t)=\left(
                                                           \begin{array}{cc}
                                                             S(j,l,t) & G(j,l,t) \\
                                                             -G(j,l,t)^T & Q(j,l,t) \\
                                                           \end{array}
                                                         \right).
\end{eqnarray}
The dimension of each block is $l-j+1$. $S(j,l,t)$ and $Q(j,l,t)$ are purely imaginary and antisymmetric, while $G(j,l,t)$ is purely real. Explicitly, the elements of the matrix are two-point correlation functions:
\begin{eqnarray}\label{twopoint}
S(j,l,t)_{mn}&=&\delta_{mn}+\langle\Psi^i_0|B_{j+m-1}(t)B_{j+n-1}(t)|\Psi^i_0\rangle,\nonumber\\
Q(j,l,t)_{mn}&=&-\delta_{mn}+\langle\Psi^i_0|A_{j+m}(t)A_{j+n}(t)|\Psi^i_0\rangle,\nonumber\\
G(j,l,t)_{mn}&=&\langle\Psi^i_0|B_{j+m-1}(t)A_{j+n}(t)|\Psi^i_0\rangle.
\end{eqnarray}
Here we use the properties of $\{A_j,A_l\}=2\delta_{jl}$, $\{B_j,B_l\}=-2\delta_{jl}$, and $\{A_j,B_l\}=0$. Finally, applying the relation between Pfaffian and determinant, we have
\begin{eqnarray}
|\langle\Psi^i_0|\mu^x_{j}(t)\mu^x_{l}(t)|\Psi^i_0\rangle|=\sqrt{\operatorname{det}\Gamma(j,l,t)}.
\end{eqnarray}

The initial state $|\Psi^i_0\rangle$ is the vacuum state of $H^i$, namely, $\eta^i_k|\Psi^i_0\rangle=0$ for every $k$. To calculate the two-point correlation function in Eq. (\ref{twopoint}), we need to express $A_l(t)$ and $B_l(t)$ by $\eta^i$ and $\eta^{i\dagger}$:
\begin{eqnarray}
\label{AtBt2eta}
A_l(t)&=\sum_k\tilde{\phi}_{lk}^\ast(t)\eta_k^{i\dagger}+\tilde{\phi}_{lk}(t)\eta^i_k,\nonumber\\
B_l(t)&=\sum_k\tilde{\psi}_{lk}^\ast(t)\eta_k^{i\dagger}-\tilde{\psi}_{lk}(t)\eta^i_k.
\end{eqnarray}
So we have
\begin{eqnarray}
\langle\Psi^i_0|\!A_m(t)\!A_n(t)\!|\Psi^i_0\rangle\!&=&\!\sum_{k}\!\tilde{\phi}_{mk}(t)\tilde{\phi}^\ast_{nk}(t)\!=\!(\!\tilde{\phi}(t)\tilde{\phi}^\dagger(t)\!)_{m\!n},\nonumber\\
\langle\Psi^i_0|\!B_m(t)\!B_n(t)\!|\Psi^i_0\rangle\!&=&\!\!-\!\sum_{k}\!\tilde{\psi}_{mk}(t)\tilde{\psi}^\ast_{nk}(t)\!=\!\!-(\!\tilde{\psi}(t)\tilde{\psi}^\dagger(t)\!)_{m\!n},\nonumber\\
\langle\Psi^i_0|\!A_m(t)\!B_n(t)\!|\Psi^i_0\rangle\!&=&\!\sum_{k}\tilde{\phi}_{mk}(t)\tilde{\psi}^\ast_{nk}(t)\!=\!(\!\tilde{\phi}(t)\tilde{\psi}^\dagger(t)\!)_{m\!n},\nonumber\\
\langle\Psi^i_0|\!B_m(t)\!A_n(t)\!|\Psi^i_0\rangle\!&=&\!\!-\!\sum_{k}\!\tilde{\psi}_{mk}(t)\tilde{\phi}^\ast_{nk}(t)\!=\!\!-(\!\tilde{\psi}(t)\tilde{\phi}^\dagger(t)\!)_{m\!n}.\nonumber
\end{eqnarray}

The matrices $\tilde{\phi}(t)$ and $\tilde{\psi}(t)$ can be expressed in a closed form.  To this end, we first consider the Heisenberg equation of the quasi-particle operator:
\begin{eqnarray}
\frac{d}{dt}\eta^f_k(t)=i[H^f,\eta^f_k(t)]=-i\omega_f\eta^f_k(t).
\end{eqnarray}
The solution is
\begin{eqnarray}
\left(
  \begin{array}{c}
    \eta^f(t) \\
    \eta^{f\dagger}(t) \\
  \end{array}
\right)
=\left(
   \begin{array}{cc}
     e^{-it\omega_f} & 0 \\
     0 & e^{it\omega_f} \\
   \end{array}
 \right)
 \left(
   \begin{array}{c}
     \eta^f \\
     \eta^{f\dagger} \\
   \end{array}
 \right).
\end{eqnarray}
Combining Eq. (\ref{c2eta}, \ref{AB}) in the Heisenberg picture, we get the linear transformation matrices in Eq. (\ref{AtBt2eta}):
\begin{eqnarray}
\tilde{\phi}(t)&=&\phi_f^T\operatorname{cos}(\omega_ft)\phi_f\phi_i^T-i\phi_f^T\operatorname{sin}(\omega_ft)\psi_f\psi_i^T,\nonumber\\
\tilde{\psi}(t)&=&\psi_f^T\operatorname{cos}(\omega_ft)\psi_f\psi_i^T-i\psi_f^T\operatorname{sin}(\omega_ft)\phi_f\phi_i^T,
\end{eqnarray}
where $\phi$ and $\psi$ are the combinations of $g$ and $h$ in Eq. (\ref{eta2c}): $\phi=g+h$, $\psi=g-h$; and their subscripts $i$ and $f$ correspond to the Hamiltonian $H^i$ and $H^f$. As a result, provided that $H^i$ and $H^f$ are numerically diagonalized, we can compute the correlation function $\langle\Psi(t)|\mu^x_j\mu^x_l|\Psi(t)\rangle$ within the numerical precision.

\section{Calculation of entanglement entropy}
The entanglement entropy is defined as $S_{A}(t)=-\operatorname{tr}[\rho_A(t)\operatorname{log_2}\rho_A(t)]$, where the subsystem consists of spins on the contiguous lattice cites $A=[1,2,\cdots,L]$. We introduce the Majorana fermions
\begin{eqnarray}
d_{2l-1}=\left(\prod_{j=1}^{l-1}\mu_{j}^{z}\right)\mu_{l}^{x},~~d_{2l}=\left(\prod_{j=1}^{l-1}\mu_{j}^{z}\right)\mu_{l}^{y}.
\end{eqnarray}
Combining Eq. (\ref{Diracfermion}) and Eq. (\ref{AB}), we have
\begin{eqnarray}
d_{2l-1}=c_{l}+c_{l}^\dagger=A_{l},~~d_{2l}=\frac{c_{l}-c_{l}^\dagger}{i}=iB_{l}.
\end{eqnarray}
The reduced density matrix can be expressed as
\begin{eqnarray}
\rho_{\!A}\!(t)\!=\!2^{-L}\!\!\!\!\!\!\!\!\!\!\!\!\!\!\!\!\!\!\!\sum_{\alpha_1,\alpha_2,\cdots,\alpha_{2L}\in\{0,1\}}\!\!\!\!\!\!\!\!\!\!\!\!\!\!\langle\!\Psi(t)|d_1^{\alpha_1}d_2^{\alpha_2}\!\cdots d_{2L}^{\alpha_{2L}}\!|\!\Psi(t)\!\rangle\left(d_1^{\alpha_1}d_2^{\alpha_2}\!\cdots d_{2L}^{\alpha_{2L}}\right)^\dagger.\nonumber
\end{eqnarray}
Notice that the fermionic parity is conserved, so if $\sum_{j=1}^{2L}\alpha_{j}=1\operatorname{mod}(2)$, then $\langle\Psi(t)|d_1^{\alpha_1}d_2^{\alpha_2}\cdots d_{2L}^{\alpha_{2L}}|\Psi(t)\rangle=0$. The none zero components can be evaluated by the Wick's theorem. It is clear that $\{d_{j},j=1,2,\cdots,2L\}$ is a orthogonal basis which spans the space of the linear operators supported on $L$. We can also find another orthogonal basis
\begin{eqnarray}\label{em}
e_m=\sum_{l=1}^{2L}V_{ml}d_l,~~~V\in O(2L),
\end{eqnarray}
to expand the reduced density matrix $\rho_L(t)$ in a direct product form.
To this end, we construct the correlation matrix
\begin{eqnarray}
\langle\Psi(t)|d_{m}d_{n}|\Psi(t)\rangle=\delta_{mn}+i\Gamma(t)_{mn},
\end{eqnarray}
where $m,n=1,2,\cdots,2L$ and
\begin{eqnarray}
\Gamma(t)_{2l\!-\!1,2s\!-\!1}&=&-i\langle\Psi^i_0|A_l(t)A_s(t)|\Psi^i_0\rangle\!=\!-i(\tilde{\phi}(t)\tilde{\phi}^\dagger(t))_{ls},\nonumber\\
\Gamma(t)_{2l,2s}&=&i\langle\Psi^i_0|B_m(t)B_n(t)|\Psi^i_0\rangle\!=\!-i(\tilde{\psi}(t)\tilde{\psi}^\dagger(t))_{mn},\nonumber\\
\Gamma(t)_{2l\!-\!1,2s}&=&\langle\Psi^i_0|A_l(t)B_s(t)|\Psi^i_0\rangle\!=\!(\tilde{\phi}(t)\tilde{\psi}^\dagger(t))_{ls},\nonumber\\
\Gamma(t)_{2l,2s\!-\!1}&=&\langle\Psi^i_0|B_l(t)A_s(t)|\Psi^i_0\rangle\!=\!-(\tilde{\psi}(t)\tilde{\phi}^\dagger(t))_{ls}.\nonumber
\end{eqnarray}
Here, $\Gamma(t)$ is a real antisymmetric matrix so it can be block diagonalized by an orthogonal matrix as
\begin{eqnarray}
V\Gamma(t)V^T=\bigoplus_{m=1}^L \nu_m(t)\left(
                                       \begin{array}{cc}
                                         0 & 1 \\
                                         -1 & 0 \\
                                       \end{array}
                                     \right),
\end{eqnarray}
where $V$ has appeared in Eq. (\ref{em}). In the new basis, the reduced density matrix is
\begin{eqnarray}
\rho_A(t)&=&\prod_{m=1}^L\frac{1}{2}\left(\langle\Psi(t)|e_{2m-1}e_{2m}|\Psi(t)\rangle e_{2m}e_{2m-1}+1\right)\nonumber\\
&=&\prod_{m=1}^L\frac{1}{2}\left(i\nu_m e_{2m}e_{2m-1}+1\right)\nonumber\\
&=&\prod_{m=1}^L\left(\frac{1-\nu_m}{2}b^\dagger_{m}b_{m}+\frac{1+\nu_m}{2}b_{m}b^\dagger_{m}\right),
\end{eqnarray}
where the Dirac fermion operators $b_m=\frac{1}{2}\left(e_{2m-1}+ie_{2m}\right)$ and $b^\dagger_m=\frac{1}{2}\left(e_{2m-1}-ie_{2m}\right)$ are introduced. we express the Von Neumann entropy as the sum of binary entropies of $L$ uncorrelated modes \cite{Latorre2003,Vidal2003},
\begin{eqnarray}
S\left(\rho_A(t)\right)=\sum_{m=1}^L H_b\left(\frac{1-\nu_m}{2}\right),
\end{eqnarray}
where
\begin{eqnarray}
H_b(x)\equiv-x\operatorname{log_2}x-(1-x)\operatorname{log_2}(1-x),
\end{eqnarray}
with $0\leq x\leq 1$, is the binary entropy.

\section{Time-evolution matrix for the ladder model}
\begin{figure}
\center
\includegraphics[width=\linewidth]{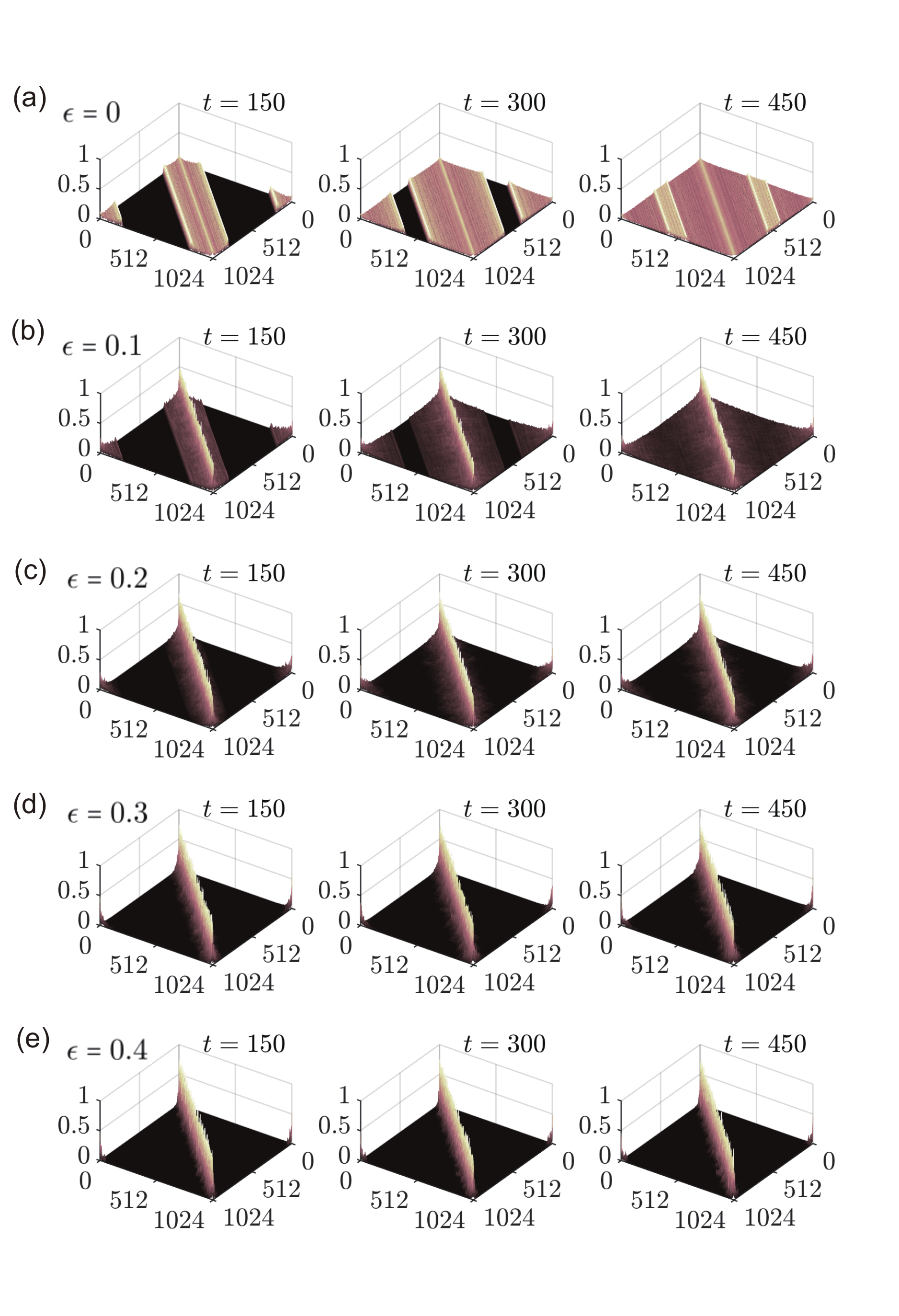}\\
\caption{Typical value of $\mathcal{M}^\prime_{i,j}(t)=\left|\mathcal{U}^\prime_{2i\!-\!1,2j\!-1\!}(t)\right|\!+\!\left|\mathcal{U}^\prime_{2i\!-\!1,2j\!}(t)\right|$ for  distinct $\epsilon$ values on short-time scales.}\label{ladderevomatshort}
\end{figure}
\begin{figure}
\center
\includegraphics[width=\linewidth]{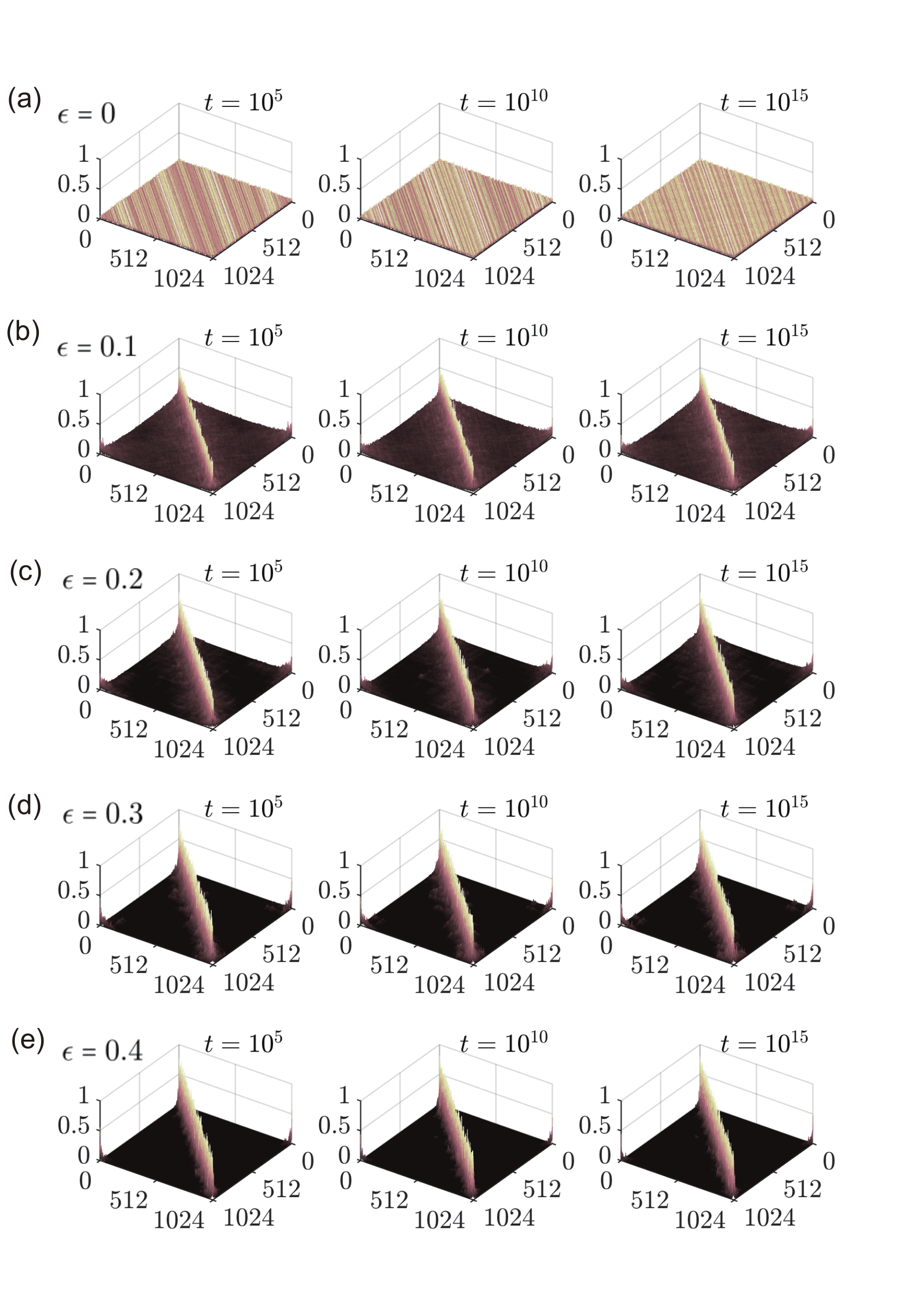}\\
\caption{Typical value of $\mathcal{M}^\prime_{i,j}(t)=\left|\mathcal{U}^\prime_{2i\!-\!1,2j\!-1\!}(t)\right|\!+\!\left|\mathcal{U}^\prime_{2i\!-\!1,2j\!}(t)\right|$ for  distinct $\epsilon$ values on long-time scales.}\label{ladderevomatlong}
\end{figure}
The Hamiltonian of the ladder model in the main text is
\begin{eqnarray}
H_{LM}=&&i\sum_{n=1}^{L}J_{n}^x d_{(n,A)}d_{(n+1,B)}\nonumber\\&-&J_{n}^y d_{(n,B)}d_{(n+1,A)}-G_nJ_{n}^zd_{(n,A)}d_{(n,B)}
\end{eqnarray}
with gauge condition $G_{n}=(-1)^n$. We assume $L=0(\operatorname{mod} 4)$, and introduce a unitary transformation:
\begin{eqnarray}
R=\prod_{k=1(\operatorname{mod} 4)}id_{(k,A)}d_{(k+1,B)}id_{(k+1,A)}d_{(k+2,B)}.
\end{eqnarray}
In what follows, we consider $H^\prime_{LM}=R^\dagger H_{LM}R$, which reads explicitly:
\begin{eqnarray}\label{HLM}
H^\prime_{LM}=&&i\sum_{n=1}^{L}J_{n}^x d_{(n,A)}d_{(n+1,B)}\nonumber\\&+&J_{n}^y d_{(n,B)}d_{(n+1,A)}-J_{n}^zd_{(n,A)}d_{(n,B)}.
\end{eqnarray}
Introduce Dirac-fermion operators $\alpha_{n}$, such that
\begin{eqnarray}
d_{(n,A)}=\frac{\alpha_{n}-\alpha_{n}^\dagger}{i},~~~~~~d_{(n,B)}=\alpha_{n}+\alpha_{n}^\dagger.
\end{eqnarray}
Eq. (\ref{HLM}) turns out to be
\begin{eqnarray}
H^\prime_{LM}&=&\sum_{n=1}^{L}-(J^x_n-J^y_n)(\alpha^\dagger_{n}\alpha_{n+1}-\alpha_{n}\alpha^\dagger_{n+1})\nonumber\\
&+&(J^x_n+J^y_n)(\alpha_{n}\alpha_{n+1}-\alpha^\dagger_{n}\alpha^\dagger_{n+1})+J^z_n(\alpha^\dagger_{n}\alpha_{n}-\alpha_{n}\alpha^\dagger_{n}).\nonumber\\
\end{eqnarray}
This is actually the fermionic representation of a 1D quantum XY model \cite{Feng2007}. The first quantized Hamiltonian is a $2\times2$-block tridiagonal Jacobi matrix (except for the boundary terms) $\mathcal{H}^\prime_{m,n}=[2J^z_m\delta_{m,n}-(J^x_{m}-J^y_{m})\delta_{m,n-1}-(J^x_{n}-J^y_{n})\delta_{m-1,n})]\sigma^z
-[(J^x_{m}+J^y_{m})\delta_{m,n-1}-(J^x_{n}+J^y_{n})\delta_{m-1,n}]i\sigma^y$, and the boundary condition is antiperiodic since the gauge condition. In the Heisenberg picture, $\alpha_i(t)\!=\!\!\sum_{j=1}^{L}\mathcal{U}^\prime_{2i-1,2j-1}(t)\alpha_{j}\!+\!\mathcal{U}^\prime_{2i-1,2j}(t)\alpha_{j}^\dagger$,
where $\mathcal{U}^\prime(t)=e^{-it\mathcal{H}^\prime}$. Setting $\{J_n^x=1+\epsilon\eta_n, J_n^y=0.1, J_n^z=0.5\}$, where $\eta_n\in[-1, 1]$ are i.i.d random variables, we calculate $\mathcal{M}^\prime_{i,j}(t)=\left|\mathcal{U}^\prime_{2i\!-\!1,2j\!-1\!}(t)\right|\!+\!\left|\mathcal{U}^\prime_{2i\!-\!1,2j\!}(t)\right|$ for distinct disorder strength $\epsilon$. The numerical results of $\mathcal{M}^\prime(t)$ for short-time scales are shown in Fig. \ref{ladderevomatshort}, and those for long-time scales are presented in Fig. \ref{ladderevomatlong}. The results show that as the disorder strength increases, the unitary time-evolution matrix tends to become closer to the identity matrix, with exponentially decaying tails.

Notice that the disorders are introduced in the subdiagonal and superdiagonal blocks of the first quantized Hamiltonian. To the best of our knowledge, there are no rigorous results about dynamical localization in this regime. However, our numerical results indicate that the dynamical-localization condition is fulfilled, resulting in the emergence of a bounded light cone.

%